\documentclass[11pt]{article}
\usepackage{moriond,epsfig}

\newcommand{\be}{\begin{equation}}
\newcommand{\ee}{\end{equation}}
\newcommand{\ba}{\begin{eqnarray}}
\newcommand{\ea}{\end{eqnarray}}
\newcommand{\bc}{\begin{center}}
\newcommand{\ec}{\end{center}}
\newcommand{\bfig}{\begin{figure}}
\newcommand{\efig}{\end{figure}}
\newcommand{\f}[2]{\frac{#1}{#2}}
\newcommand{\g}{\gamma}
\newcommand{\om}{\omega}

\newcommand{\al}{\alpha}
\newcommand{\bal}{\bar \alpha}

\newcommand{\rr}[4]{#1, {\it #2 \/}{\bf #3} #4}

\begin{document}
\vspace*{4cm}
\title{Phenomenology of next-leading BFKL } 

\author{ L. SCHOEFFEL }

\address{CEA/DSM/DAPNIA/SPP, F-91191 
Gif-sur-Yvette Cedex}

\maketitle\abstracts{
We propose a phenomenological study of the Balitsky-Fadin-Kuraev-Lipatov
(BFKL) approach applied to the data on  the proton structure 
function $F_2$ measured at HERA in the small-$x_{Bj}$ region
($x_{Bj}<0.01$) and $Q^2$ in the range $5$ to $120$~GeV$^2$. With a 
simplified ``effective kernel'' approximation, we present   a  comparison between 
leading-logs (LO) and next-to-leading logs (NLO) BFKL approaches in the 
saddle-point approximation, using known resummed NLO-BFKL kernels. 
The LO result gives a very good description of the data with a three parameters fit of $F_2$ but
an unphysical  value of the  strong coupling constant,
whereas the NLO two parameters fit leads to a   qualitatively satisfactory account of the running  
coupling constant effect  but quantitatively, for $Q^2 < 10$~GeV$^2$, it
fails to reproduce properly the data.
}

\section{Introduction}

Precise measurements of the proton structure function $F_2$ 
at small $x_{Bj}$ ($x_{Bj}<0.01$ )~\cite{data} have led to
important tests of QCD evolution equations
and a better understanding
of deep inelastic scattering phenomenology. For  the 
Dokshitzer-Gribov-Lipatov-Altarelli-Parisi~\cite{dglap}
(DGLAP) evolution in  $Q^2$, it has  been possible 
to test it in various ways with  NLO (next-to-leading $\log Q^2$) and 
now NNLO  accuracy and it works quite well in a large range of $Q^2$ and $x_{Bj}.$
Testing precisely the Balitsky-Fadin-Kuraev-Lipatov~\cite{bfkl}
(BFKL) evolution in $x_{Bj}$ beyond leading order appears 
more difficult and it is the aim of the follwoing analysis.

In a first part, we perform a LO-BFKL analysis of $F_2$ measurements~\cite{data}, 
reproducing results obtained previously with
older data~\cite{old}. However, as it was found in reference~\cite{old},
we have to introduce an effective but unphysical  value of the  strong coupling constant  
$\al \sim 0.1$ instead of $\al \sim 0.2$ predicted for the $Q^2$-range considered here.
It reveals the need for NLO corrections. 

In fact,  the  theoretical task of computing these   
corrections appears to be quite hard. It is now in good progress but  
still under completion for the coupling to external particles. For the BFKL kernel, they have been 
calculated after much efforts~\cite{next}. 
In addition, for the whole theoretical
approach to be correct, an appropriate resummation of spurious singularities, 
brought together with the 
NLO corrections, has to be performed at all orders of the perturbative 
expansion~\cite{salam}.
This resummation procedure is
required by consistency with  the QCD renormalization group. 
Various resummation 
schemes have been proposed~\cite{salam,autres,lipatov} which satisfy the 
renormalization group requirements while retaining the computed  value of the 
NLO terms  in  the BFKL kernel. 
In the following, we present an NLO-BFKL analysis for the
scheme labeled $S3$ in reference~\cite{salam}. 
A more complete treatment can be found in reference~\cite{robinlo}, where
we compare the different NLO schemes  in order  to  distinguish 
between different resummation options.

%%%%%%%%%%%%%%%%%%%%%%%%%%%%%%%%%%%%%%%%%%%%%%%%%%%%%%%%%%%%%%%%%%%%%%%%
%%%%
%%%%
%%%%

\section{``Effective kernel'' and saddle-point approximation of BFKL 
amplitudes}
\label{2}

The BFKL formulation of the proton structure functions  can be formulated in 
terms of  the double inverse Mellin integral
\begin{equation}
{F_2}=\int \! \int 
\frac {d\gamma d\om}{(2i\pi)^2} \left(\frac{Q^2}{Q_0^2}\right)^{\gamma} 
{x_{Bj}}^{-\om} \   {{\cal F}_2}(\g,\om)\ .
\label{bfkl0}
\end{equation}

At LO level one has (see, e.g.~\cite{old}) 
\begin{equation}
{{\cal F}_2}(\g,\om)=\f {h_2(\g,\om)}{\om-\bal \ 
\chi_{L0}(\gamma)}
\label{bfkl}
\end{equation}
where $\bal \equiv  {\al_s N_c}/{\pi},$   $\al_s$ is the coupling 
constant which is merely  a parameter
at this LO level. The LO BFKL kernel is written as
\begin{equation}
\chi_{LO}(\gamma)=2\psi(1)-\psi(\gamma)-\psi(1-\gamma)\ .
\label{kernel}
\end{equation}
$h_2(\g,\om)$ is a prefactor  which  takes into account  both the  
phenomenological non-perturbative coupling to the proton and the   perturbative coupling 
to the virtual photon. Note that the  variable $\gamma$  plays the role of a continuous 
anomalous dimension while $\om$ is the continuous index of the Mellin moment 
conjugated with the rapidity $Y\equiv \log {1}/{x_{Bj}}$. 

\begin{figure}[!]
\begin{center}
\epsfig{figure=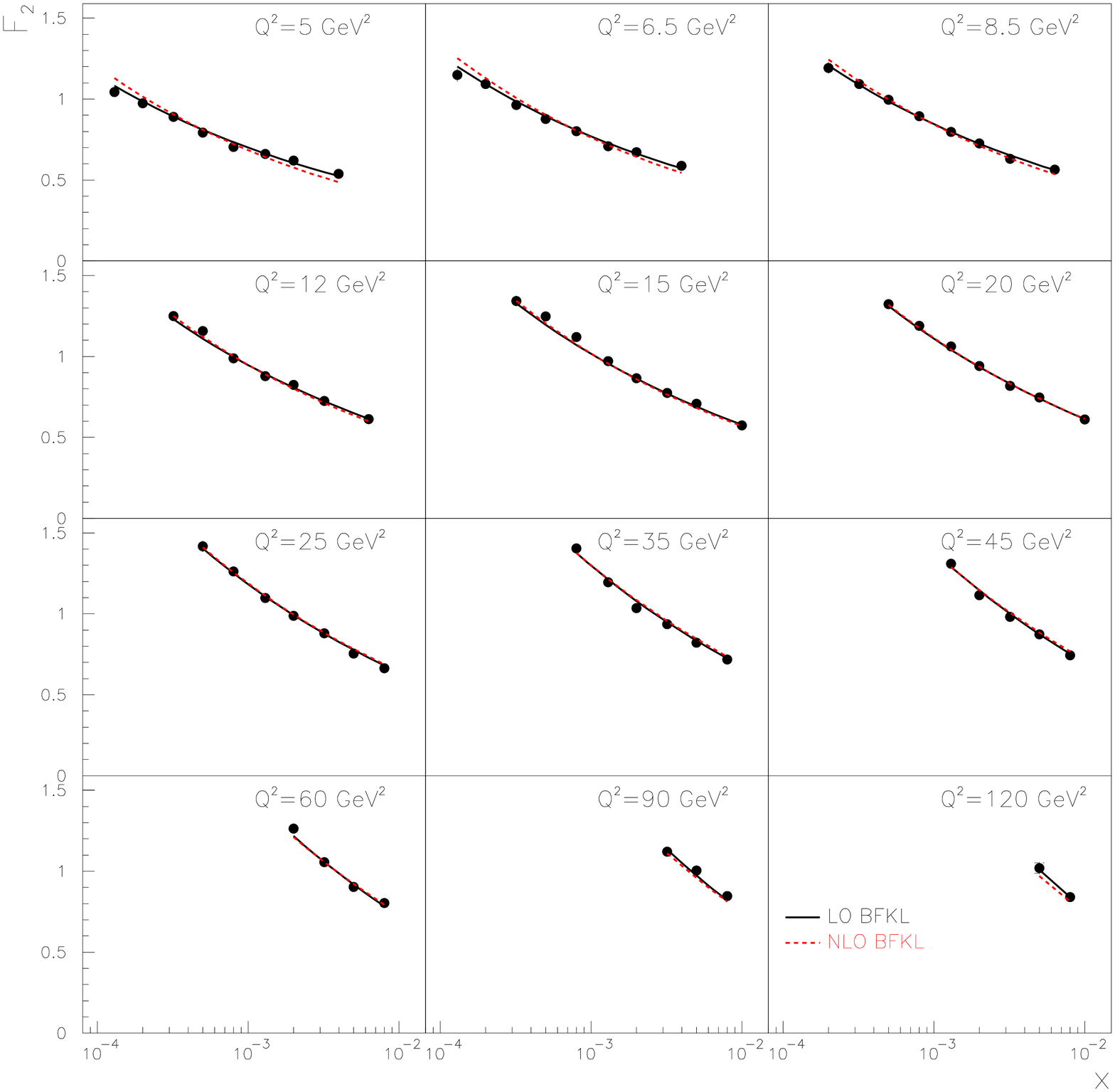,height=4.in}
\end{center}
\caption{Results of the BFKL fits to the H1 data:
LO (continuous lines) and NLO -scheme $S3$- (dashed lines).
 }
\label{fig_fitx_2}
\end{figure}

Recalling  well-known properties of LO-BFKL amplitudes, one  assumes that 
$h_2(\g,\om)$ is regular. The  pole contribution  at
$\om = \bal \ \chi_{L0}(\gamma)$
in  (\ref{bfkl}) leads  to a single Mellin transform in $\g$ for which one 
may use a saddle-point approximation 
at  small values of ${x_{Bj}}$,
which is known to give a very good account of the phenomenology at LO
\begin{equation}
{F_2(x,Q^2)}\approx \ {\cal N}
\ 
\exp {\left\{\frac L2 + {\al_s} Y \chi_{L0}({\scriptstyle \frac 12}) -  
\frac 
{L^2}{2{\al_s} 
Y \chi''_{L0}(\frac 12)}\right\}}\ ,
\label{approxbfkl}
\end{equation}
where  $L\equiv 
\log({Q^2}/{Q_0^2})$ and ${\cal N}$ is a normalisation taking into account 
all the smooth prefactors\footnote{In particular, the square root prefactor   of 
the gaussian saddle-point 
approximation can be  merged in the normalization.}.  As a 
consequence, the only 
three relevant parameters in (\ref{approxbfkl}) are ${\cal N}, \bal$ and 
$Q_0.$ In this picture $\bal$ has to be considered as a
parameter and not a genuine  QCD coupling constant since the value obtained  
in the  fits  is not related to the coupling constant values in the 
considered  range of $Q^2$. The result of the fit is presented on figure~\ref{fig_fitx_2}
together with the measurements of $F_2$ by the H1 collaboration~\cite{data}. 
We obtain a very good agreement (at the $\chi^2$ level of the NLO DGLAP fit) between 
this simple parameterisation and $F_2$ data.

\begin{figure}[!]
\begin{center}
\epsfig{figure=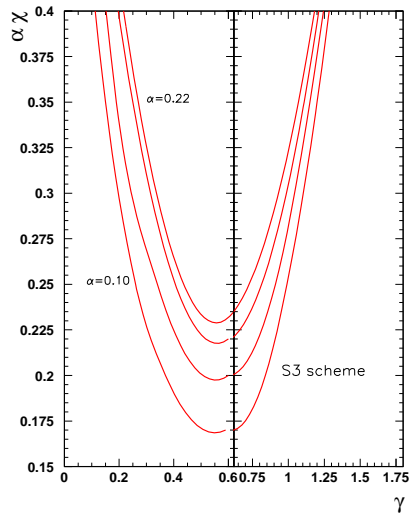,height=3.in}
\end{center}
\caption{$\alpha_{RG} \chi$ as a function of $\gamma$ for different values of
$\al_{RG}$  for $S3$ scheme
($\alpha_{RG}$ 
varies 
between 0.1 to 0.24). 
 }
\label{fig_fitx_1}
\end{figure}

The treatment of the NLO BFKL kernel (expressed with the resummation scheme $S3$)
is detailed in reference~\cite{robinlo}.
It retains the  running 
property of the QCD coupling constant with its theoretically predetermined value at the 
relevant $Q^2$ range. 
It is shown that an "effective kernel" approximation associated with
a saddle-point expression similar to (\ref{approxbfkl}) allows one to obtain
a simple formula for the structure function $F_2$

\begin{equation}
{F_2(x,Q^2)}\approx \ {\cal N}
\ 
\exp {\left\{
\g_c \ L + {\al_{RG}}\ 
\chi_{eff}(\g_c,\al_{RG})\ Y  - 
\frac {L^2}{2\al_{RG}\chi''_{eff}(\g_c,\al_{RG})\ Y} \right\}}\ ,
\label{approxnlo}
\end{equation}
where $\chi_{eff}$ is the "effective kernel" at NLO which is shown in figure~\ref{fig_fitx_1}
and $\g_c$ is defined by the implicit  saddle-point equation 
\be
\frac{\partial \chi_{eff}}{\partial \g}(\g_c,\al_{RG}(Q^2))= 0\ \ \ \ \ \ \
\left[\al_{RG}(Q^2)\right]^{-1} \equiv {b\
\log\left(Q^2/\Lambda_{QCD}^2\right)}.
\label{gc2}
\ee
with $b=11/12-1/6\ N_f/N_c$.
It is important at this stage to notice that the formula  
(\ref{approxnlo}) has only two free parameters 
${\cal N}$ and $Q_0$ instead of three for the LO case (\ref{approxbfkl}), as we are using 
the QCD universal coupling constant ${\al_{RG}}.$ It allows one to compare in a similar footing 
the LO and  NLO BFKL kernels to    $F_2$ 
data, as presented on figure~\ref{fig_fitx_2}. A global good agreement is obtained
but the for $Q^2 < 10$~GeV$^2$, we notice that the NLO BFKL fit 
fails to reproduce properly the $F_2$ data. A similar conclusion holds for other aviable
resummation schemes~\cite{robinlo}.

%%%%%%%%%%%%%%%%%%%%%%%%%%%%%%%%%%%%%%%%%%%%%%%%%%%%%%%%%%%%

\section{Conclusion}
\label{5}
We have confronted the predictions of BFKL kernels 
at the level of leading and  next-leading  logarithms (scheme $S3$) with structure 
function data. We have proposed to use the ``effective kernel'' 
approximation of the NLO-BFKL kernels which, associated with the usual  
saddle-point approximation at high rapidity and large enough  $Q^2$, allows 
one to obtain a simple two parameters formula for the structure function 
$F_2$. The comparison with the similar three parameters formula commonly used at 
LO level shows a deterioration of the fits for $Q^2 < 10$~GeV$^2$.
These deviations are under study and could originate from different sources as
the non validity of the saddle-point approximation at low $Q^2$ or unknown aspects of the 
prefactors, in particular the non-perturbative ones.
 
\section*{Acknowledgments}
I thank R. Peschanski and C. Royon for a fruitful
collaboration on this subject.

\end{document}